\documentclass[aps,prl,twocolumn,superscriptaddress,showpacs,amssymb,amsmath]{revtex4-1}

\usepackage{graphicx}

\usepackage{units}
\usepackage{txfonts}

\begin{document}
\title{Spin relaxation near the metal-insulator transition: dominance of the 
Dresselhaus spin-orbit coupling}

\author{Guido A.\ Intronati}
\author{Pablo I.\ Tamborenea}
\affiliation{Departamento de F\'{\i}sica and IFIBA, FCEN, 
Universidad de Buenos Aires, Ciudad Universitaria, Pab.\ I, 
C1428EHA Ciudad de Buenos Aires, Argentina}
\affiliation{Institut de Physique et Chimie des Mat\'{e}riaux de Strasbourg, 
UMR 7504, CNRS-UdS, 23 rue du Loess, BP 43, 67034 Strasbourg Cedex 2, France}

\author{Dietmar Weinmann}
\author{Rodolfo A.\ Jalabert}
\affiliation{Institut de Physique et Chimie des Mat\'{e}riaux de Strasbourg, 
UMR 7504, CNRS-UdS, 23 rue du Loess, BP 43, 67034 Strasbourg Cedex 2, France}

\begin{abstract}
We identify the Dresselhaus spin-orbit coupling as the source of the dominant
spin-relaxation mechanism in the impurity band of doped semiconductors.
The Dresselhaus-type (i.e.\ allowed by bulk-inversion asymmetry) hopping terms are derived 
and incorporated into a tight-binding model of impurity sites, and they are shown to 
unexpectedly dominate the spin relaxation, leading to spin-relaxation times 
in good agreement with experimental values.
This conclusion is drawn from two complementary approaches employed to extract the
spin-relaxation time from the effective Hamiltonian: an analytical
diffusive-evolution calculation and a numerical finite-size scaling.
\end{abstract}

\maketitle

Spin dynamics in semiconductors is a fundamental issue in view of the rich
physics involved and the potential technological applications \cite{fab-etal,
wu-jia-wen}.
It is thus not surprising that spin relaxation studies were already performed
in the early days of semiconductor research \cite{ell,yaf,ale-hol} and are
intensely pursued today with modern experimental techniques \cite{kik-aws,
oes-roe-hau-hae}.
An intriguing experimental observation is the fact that in n-doped semiconductors 
at low temperatures the spin relaxation time $\tau_\mathrm{s}$ presents a maximum 
as a function of the doping density near the metal-insulator 
transition (MIT) \cite{ale-hol,cha,zar-cas,kik-aws,dzh,sch-hei-roh,roe-ber-mue}. 

Interestingly, while the mechanisms behind spin relaxation have been properly 
identified at high temperatures or for doping densities away from the critical 
one \cite{dzh,lau01,jia-wu}, a theoretical understanding of low-temperature 
spin relaxation close to the MIT is still lacking.
This unsatisfactory state of affairs has motivated some attempts to 
identify the relevant mechanisms for spin relaxation 
\cite{shk,kav,put-joy,tam-wei-jal} close to the MIT.
In particular, on the metallic side of the transition, Shklovskii has proposed the 
applicability of the well-known Dyakonov-Perel mechanism usually valid
in the conduction band \cite{shk}.
Furthermore, a tight-binding model of impurities including Rashba
spin-orbit coupling has been developed \cite{tam-wei-jal}.
The spin relaxation times resulting from this last model were larger than
the experimental values, implying that other mechanisms should be active
in this density regime.

In this work we identify the Dresselhaus coupling as the source of the
leading spin relaxation mechanism on the metallic side of the transition.
This conclusion is based on the construction of an effective spin-orbit 
Hamiltonian for the impurity system of III-V semiconductors, together with its 
analytical and numerical solution.
The resulting spin relaxation times are in good agreement with the existing
experimental values for GaAs.
The detailed temperature-dependent measurements of Ref.~\cite{roe-ber-mue} 
yielded a saturation of $\tau_\mathrm{s}$ below \unit[10]{K}, indicating that inelastic 
processes are irrelevant at low temperatures. 
We thus work with a zero-temperature formalism.

The envelope-function approximation (EFA) for describing conduction-band
electrons in zincblende semiconductors incorporates the lattice-scale
physics (described by the periodic part of the Bloch wave function) into
the effective one-body Hamiltonian \cite{noz-lew,eng-ras-hal}
\begin{eqnarray} 
H &=& H_0 + H_{\text{SIA}} + H_{\text{BIA}}
\label{eq:Htot}
\\ 
H_0 &=& \frac{p^2}{2 \, m^*} + V(\mathbf{r})
\label{eq:Hzero}
\\ 
H_{\text{SIA}} &=& \lambda \, \boldsymbol{\sigma} \cdot \nabla V 
                 \times \mathbf{k}
\label{eq:SIA}
\\
H_{\text{BIA}} &=& \gamma \, [\sigma_x k_x (k_y^2 - k_z^2) + 
                          \text{cyclic permutations}].  
\label{eq:BIA}
\end{eqnarray}
The electrostatic potential $V(\mathbf{r})$ includes all potentials
aside from the crystal one, while $\boldsymbol{\sigma}$ is the vector of
Pauli matrices and $\mathbf{k} = \mathbf{p}/\hbar$.
The effective spin-orbit coupling $\lambda$, enabled by the structural
inversion asymmetry (SIA) is usually orders of magnitude larger than the
one of vacuum $\lambda_0 = \hbar^2 / 4 \, m_0^2 c^2 \simeq 
\unit[3.7 \times 10^{-6}]{\AA^2}$. 
It can be calculated at the level of the 8-band Kane model,
which, for example, for GaAs yields
$\lambda \simeq \unit[-5.3]{\AA^2}$ \cite{eng-ras-hal}.
The bulk inversion asymmetry (BIA) coupling constant $\gamma$ is another 
material-dependent parameter.
The exact value of $\gamma$ is a matter of current debate \cite{eng-ras-hal,fab-etal}.
A 14-band model is required for the theoretical estimation of $\gamma$,
leading to $\gamma \approx \unit[27]{eV\AA^3}$ for GaAs \cite{eng-ras-hal,win}.
More refined theoretical calculations yield somewhat lower values
\cite{car-chr-fas,cha-van-kot,kri-hal}.
While early experimental values obtained in bulk samples agree approximately
with the above-quoted value of $\unit[27]{eV\AA^3}$
\cite{mar-ste-tit}, recent results inferred from measurements in 
low-dimensional systems are again consistently lower 
\cite{knap-etal,mei-etal,fab-etal,jusserand}.

In order to study the spin relaxation in the impurity band near the
MIT, we consider the potential $V(\mathbf{r})$ due to the ionized
impurities, given by
\begin{equation} 
V(\mathbf{r}) = \sum_p V_p(\mathbf{r}) = 
                - \sum_p \frac{e^2}{\epsilon |\mathbf{r}-\mathbf{R}_p|},
\end{equation}
where $\epsilon$ is the dielectric constant of the semiconductor and 
$\mathbf{R}_p$ represents the impurity positions.
The potential $V_p$ gives rise to hydrogenic states centered at the
impurity $p$.
In order to build the basis of electronic states we will only consider 
the ground state $\phi_p(\mathbf{r})= \phi(|\mathbf{r}-\mathbf{R}_p|)$, with 
$\phi(\mathbf{r}) = (1/\sqrt{\pi a^3}) \exp{(-r / a)}$,
and $a$ the effective Bohr radius.

The second-quantized form of the Hamiltonian (\ref{eq:Htot}), 
that we denote $\mathcal{H}$, has components
\begin{eqnarray}
\label{eq:MT}
\mathcal{H}_0 &=& \sum_{m\neq m',\sigma} 
                \langle m' \sigma | H_0 | m \sigma\rangle \
                 c_{m' \sigma}^{\dag} \ c_{m \sigma}^{\phantom{\dag}}\, ,
\\
\label{eq:MTSO}
\mathcal{H}_{\text{SO}} &=& \sum_{m\neq m',\sigma} 
                \langle m' \overline{\sigma} | H_{\text{SO}} |m\sigma\rangle \
                 c_{m' \overline{\sigma}}^{\dag} \ c_{m \sigma}^{\phantom{\dag}}\, ,
\end{eqnarray}
where the label SO stands for SIA or BIA.
We denote the $1s$ state $\phi_m(\mathbf{r})$ with spin $\sigma=\pm 1$ in 
the z-direction by $|m \sigma \rangle$, and $c_{m \sigma}^{\dag}$ 
($c_{m \sigma}$) is the creation (annihilation) 
operator of a particle in that state ($\overline{\sigma}=-\sigma$). 
The matrix elements in Eq.\ (\ref{eq:MT}) contain three-center integrals
$\langle m' \sigma | V_p | m \sigma\rangle$ with $p \neq m$.
Due to the exponential decay of $\phi_m(\mathbf{r})$, one usually keeps
only the term 
\begin{eqnarray} \label{eq:t_MT}
\langle m'\sigma | V_{m'} | m \sigma \rangle 
= -V_0 \left(1+\frac{R_{m'm}}{a} \right)
    e^{- R_{m'm} / a},
\end{eqnarray}
with $V_0=e^2 /\varepsilon a$ (twice the binding energy of an isolated
impurity) and $R_{m'm}$ the distance between the two impurities.
The resulting Hamiltonian $\mathcal{H}_0$ defines the well-known 
Matsubara-Toyozawa model (MT) \cite{mat-toy}, which has been thoroughly 
studied in the context of the MIT.
The subtleties, drawbacks and applicability of this model to describe the 
metallic side of the MIT, as well as its extension to include the 
$\mathcal{H}_{\text{SIA}}$ spin-orbit coupling, have recently been discussed 
\cite{proceedings}. 
Electron-electron interactions induce significant many-body effects 
on the insulating side of the transition, but not on the metallic side. 
Therefore we do not need to include them in our model. 
According to the Mott criterion the critical dimensionless impurity density 
for the MIT corresponds to $\mathcal{N}_\mathrm{i}=n_\mathrm{i}a^3\simeq 0.017$,
corresponding to a critical density of $\unit[2\times 10^{16}]{cm^{-3}}$ for GaAs. 

The matrix element of $\mathcal{H}_{\text{SIA}}$ is
\begin{eqnarray} 
 &&\langle m' \overline{\sigma} | H_{\text{SIA}}|m\sigma\rangle =
\frac{\sigma \lambda}{a^2} \int d\mathbf{r} \,  
  V(\mathbf{r}) \,
  \frac{\phi_{m'}(\mathbf{r}) \,\phi_m(\mathbf{r})}
       {|\mathbf{r} - \mathbf{R}_{m'}| |\mathbf{r} - \mathbf{R}_m|} \nonumber \\
  &&\times \ [(z-z_{m})(r_{\sigma}-R_{m'\sigma})
           -(z-z_{m'})(r_{\sigma}-R_{m\sigma})],
\label{eq:SIA_me2}
\end{eqnarray}
where $r_{\sigma}=x+i\sigma y$ and $R_{m\sigma}=X_m+i\sigma Y_m$.
The Hamiltonian $\mathcal{H}_{\text{SIA}}$ represents the generalization 
of the Rashba coupling to the case of impurity potentials, and was 
introduced in Ref.~\cite{tam-wei-jal}.
There, an alternative path to the EFA was followed in order to calculate the 
matrix elements $\langle m' \overline{\sigma} | H_{\text{SIA}}|m\sigma\rangle$,
which made use of impurity states with spin admixture obtained 
from spin-admixed conduction-band Bloch states derived at the 
level of the 8-band Kane model.
We remark that the terms corresponding to $p=m, m'$ in $V(\mathbf{r})$
give vanishing contributions to the SIA matrix element due to the axial
symmetry of the two-center integrals.
As a consequence, the Rashba matrix elements are given by three-center
integrals, resulting in very slow spin relaxation \cite{tam-wei-jal} in 
comparison with experimental results. 
We therefore turn to the Dresselhaus term, whose matrix element 
is given by
\begin{eqnarray} 
\label{eq:matelbia}
\langle m' \bar{\sigma} | H_{\text{BIA}} | m \sigma \rangle &=&
   \gamma \,
   [\langle m' |k_x (k_y^2 - k_z^2) | m \rangle  \nonumber \\
   && +  i \, \sigma 
    \langle m' | k_y (k_z^2 - k_x^2) | m \rangle] \nonumber \\
   &=& \frac{\gamma}{\pi a^3}
      \left( \sigma \, I_{y,m'm} + i \, I_{x,m'm} \right), 
\end{eqnarray}
where
\begin{eqnarray} 
I_{x,m'm} &=&
    \frac{1}{a^3}   
    \int d\mathbf{r} \,
    \frac{e^{-|\mathbf{r}-\mathbf{R}_{m'm}|/a} \, e^{-r/a}}
         {|\mathbf{r} - \mathbf{R}_{m'm}| \, r^3} \nonumber \\
    && \times \ (a+r) (x-X_{m'm}) (y^2-z^2), 
\end{eqnarray}
and $I_{y,m'm}$ is obtained from $I_{x,m'm}$ with the exchanges
$X_{m'm} \leftrightarrow Y_{m'm}$ and $x \leftrightarrow y$.
Performing a rotation of the coordinate system and switching to
cylindrical coordinates allows us to easily do the angular integral,
yielding
\begin{eqnarray} 
\label{eq:Ixmmp}
 && I_{x,m'm} = \frac{\pi c}{a^3} 
              \int d\rho \, dz \, \rho \,
            \frac{ e^{-\sqrt{\rho^2+(z-R_0)^2}/a} \, e^{-\sqrt{\rho^2+z^2}/a}}
                 { \sqrt{\rho^2+(z-R_0)^2} \, (\rho^2+z^2)^{3/2}} \nonumber \\
            && \times \ \left(a+\sqrt{\rho^2+z^2}\right) 
                \left[ \frac{\rho^2}{2} (3z-R_0) - z^2 (z-R_0)\right],  
\end{eqnarray}
where
\begin{equation}
  c = 2 \cos\varphi \sin\theta \, [1 - \sin^2 \theta \, (1+\sin^2\varphi)],
\end{equation}
and $(R_0,\theta,\varphi)$ are the polar coordinates of $\mathbf{R}_{m'm}$
in the original reference frame.
The integrals in (\ref{eq:Ixmmp}) are not analytically solvable, 
but they can easily be integrated numerically. 
This is the route that we take later, where we simulate the dynamical evolution 
of initially prepared pure spin states. 
Before tackling the numerical problem, we provide a simple estimation 
of the spin lifetime based on the steepest-descent approximation of the integral 
\eqref{eq:Ixmmp} (valid in the limit $R_0 \gg a$), given by 
\begin{equation} 
\label{eq:Ixmmpsd}
I_{x,m'm} \simeq \frac{\pi^2 c}{4 a^2} \ R_0 \left(a+\frac{R_0}{2}\right) 
                e^{-R_0/a} \, .  
\end{equation}
The spatial diffusion of electrons through the network of impurities is
accompanied by a spin diffusion characterized by a typical spin rotation
angle per hop \cite{tam-wei-jal}
\begin{equation}\label{eq:tyrotan}
\langle \Theta^2 \rangle = \frac{15}{2} \ 
       \frac{\sum_{m\neq m'} 
       |\langle m' \overline{\sigma} | H_{\text{BIA}} |m\sigma\rangle|^2}
       {\sum_{m\neq m'} 
       |\langle m' \sigma | H_0 | m \sigma\rangle|^2} \, .
\end{equation}
The spin-relaxation time can be defined as the time in which the 
accumulated rotation reaches a unit angle, and therefore is given by
\begin{equation}\label{eq:reltimedef}
\frac{1}{\tau_\mathrm{s}} = \frac{2}{3} \frac{\langle \Theta^2 \rangle}{\tau_\mathrm{c}} \ ,
\end{equation}
where $\tau_\mathrm{c}$ is the typical hopping time.
It can be shown that this expression is independent of whether the
initial state is localized or extended \cite{tam-wei-jal}. 
Taking $\tau_\mathrm{c}$ as the time needed for the initial-state population 
on an impurity site to drop from 1 to 1/2, we obtain
\begin{equation}\label{eq:tauc}
\frac{1}{\tau_\mathrm{c}}= \frac{\sqrt{2}}{\hbar} \left( \sum_{m\neq m'} 
       |\langle m' \sigma | H_0 | m \sigma\rangle|^2\right)^{1/2} 
\simeq \frac{\sqrt{14\pi}V_0}{\hbar} \mathcal{N}_\mathrm{i}^{1/2}\ .
\end{equation}
For the second equality \cite{note-tauc} we have used the impurity average assuming
a random distribution without hard-core repulsive effects on the scale
of the effective Bohr radius \cite{thom-rice,tam-wei-jal}. 
Performing also the impurity average in Eq.\ \eqref{eq:tyrotan} we have
\begin{equation}\label{eq:tyrotanimpav}
\langle \Theta^2 \rangle = 0.258 \ 
       \left(\frac{\gamma}{a^3V_0}\right)^2 \ ,
\end{equation}
and
\begin{equation}\label{eq:taus_vs_dens}
\frac{1}{\tau_\mathrm{s}} = \frac{1.14 \ \gamma^2}{a^6V_0\hbar} \, 
\mathcal{N}_\mathrm{i}^{1/2}. 
\end{equation}
This expression for the spin relaxation time is our main result. 
As we discuss below, its confrontation with the experimental data 
allows us to identify the Dresselhaus coupling as the dominant spin relaxation 
mechanism in the regime of study. 
Furthermore, numerical calculations of 
the relaxation time within our model provide a complementary path 
validating the analytical approach, since some of the previously used 
approximations can be avoided. 

\begin{figure}
\includegraphics[width=\linewidth]{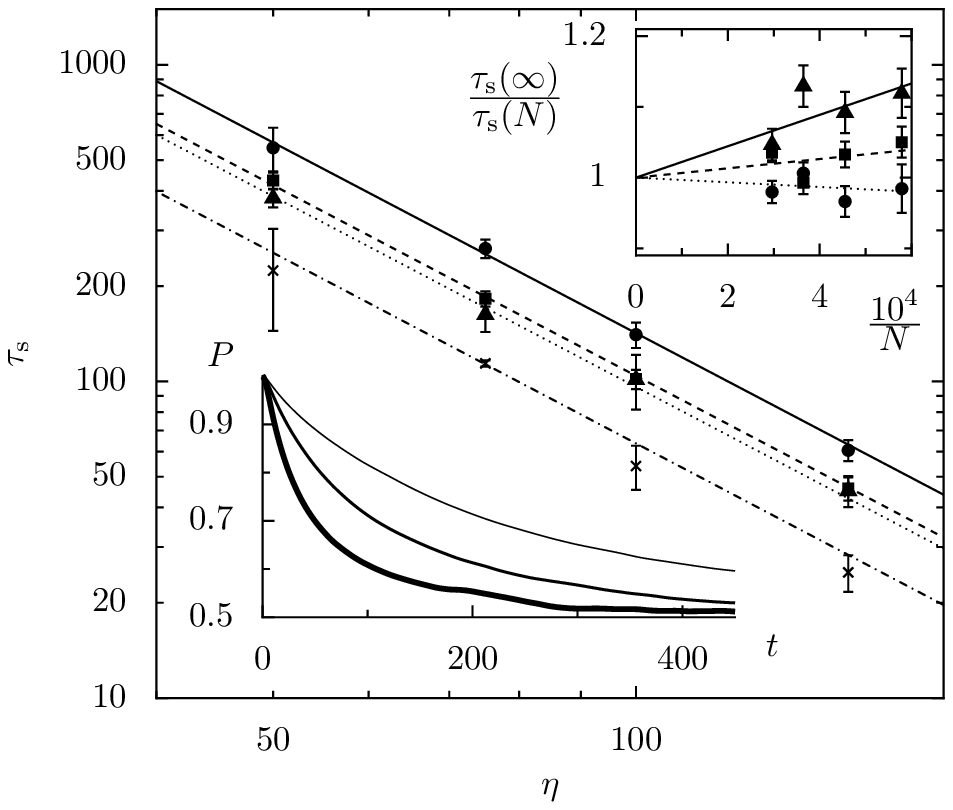}
\caption{Scaling of the spin relaxation time 
extrapolated to infinite system size, with the spin-orbit enhancement 
factor $\eta$, for densities $\mathcal{N}_\mathrm{i}=0.02$ (circles), 0.029 (squares), 
0.037 (triangles), and 0.06 (crosses). 
The lines are fits of a quadratic dependence of
the relaxation rate on $\eta$. 
The times are given in units of $\hbar/V_0$.
\textit{Lower Inset:} Spin survival probability $P$ for an initial 
MT eigenstate evolving under the enhanced Dresselhaus couplings 
for a density $\mathcal{N}_\mathrm{i}=0.029$ and a system size $N=3375$. 
Lines of increasing thickness are for enhancement factors 
$\eta=75$, 100, and 150. 
\textit{Upper Inset:} Size dependence of $\tau_\mathrm{s}^{-1}$ for $\eta=150$. 
Lines are linear fits to the data that allow to extrapolate to the infinite-size 
values.}
\label{fig:dynamevol}
\end{figure}

The numerical procedure starts from the numerical integration of 
\eqref{eq:Ixmmp} for a given impurity configuration in order to obtain the 
matrix elements \eqref{eq:matelbia} for $H_\mathrm{BIA}$ (and similarly for 
$H_\mathrm{SIA}$). 
We then diagonalize the total Hamiltonian $\mathcal{H}$ including the two 
contributions to the spin-orbit coupling, which allows us to obtain the quantum 
evolution of an arbitrary state. 
Choosing an initial state with a well-defined 
spin projection (for instance a MT eigenstate with $\sigma=1$) 
we can follow the spin evolution and extract the spin lifetime from it. 
The weakness of the spin-orbit coupling translates into spin-admixture 
perturbation energies which are, even for the largest system sizes (in terms 
of number of impurities, $N$) that we are able to treat numerically, orders of 
magnitude smaller than the typical MT level spacing. 
This large difference between the two energy scales in finite size 
simulations masks the spin-orbit-driven physics, and forces us to follow an 
indirect path: we introduce an artificially enhanced 
coupling constant $\eta \gamma$ and a finite-size scaling procedure.
The limits $N \rightarrow \infty$  and then $\eta \rightarrow 1$ taken
at the end of the calculation provide the sought spin-relaxation rate.  
The numerically extracted values of the spin relaxation times associated with 
$H_{\mathrm{SIA}}$ are, consistently with the results of 
Ref.~\cite{tam-wei-jal}, considerably larger than the ones experimentally 
observed. 
Therefore, we hereafter neglect the Rashba term in the numerical 
calculations, and concentrate on the spin evolution governed by $H_{\mathrm{BIA}}$.

\begin{figure}
\includegraphics[width=\linewidth]{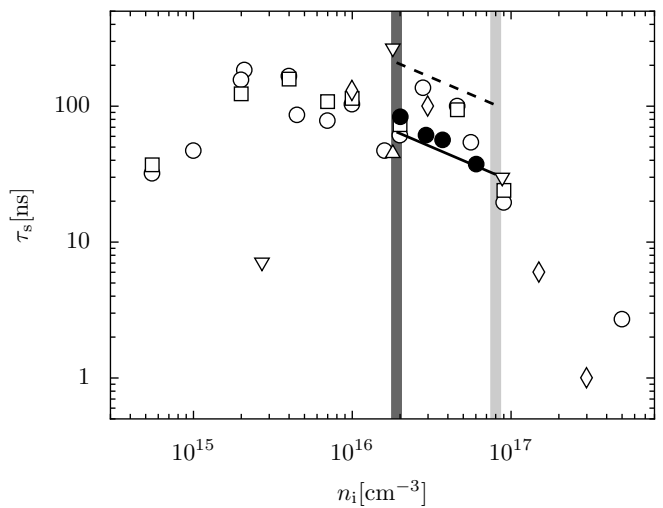}
\caption{Spin relaxation time in n-doped GaAs as a function of the doping
density. 
The prediction of Eq.\ \eqref{eq:taus_vs_dens} (solid line) and our numerical
results (filled circles) for the regime between the metal-insulator transition 
(dark gray line) and the hybridization of the impurity band with the conduction 
band (light gray line) obtained using $\gamma=\unit[27]{eV\AA^3}$ are compared 
to experimentally obtained values (open symbols) using different methods. 
Circles and squares are from 
Ref.\ \cite{dzh} for $T=\unit[2]{K}$ and $T=\unit[4.2]{K}$, respectively, 
along with data from Ref.\ \cite{kik-aws} (diamonds), 
\cite{oes-roe-hau-hae} (triangles), and \cite{roe-ber-mue} 
(reversed triangles). 
The dashed line represents the result of 
Eq.\ \eqref{eq:taus_vs_dens} with $\gamma=\unit[15]{eV\AA^3}$.
}
\label{fig:relaxtime}
\end{figure}
In the lower inset of Fig.~\ref{fig:dynamevol} we show typical spin evolutions starting 
from an eigenstate of the MT system with $\sigma=1$ in the energy range of
extended states of the impurity band, for a density
$\mathcal{N}_\mathrm{i}=0.029$ just above the MIT transition for three values of
the coupling constant and $N=3375$ impurities. 
The initial perturbative regime with a quadratic time decay of the spin
survival is followed by an exponential decay from which the relaxation rate 
$\tau_\mathrm{s}^{-1}$ can be inferred, until the saturation value of $1/2$.
For each density and effective coupling constant $\eta\gamma$ the asymptotic 
value of $\tau_\mathrm{s}^{-1}$ can be obtained by extrapolating the finite-$N$ 
values to the infinite size limit (upper inset of Fig.\ \ref{fig:dynamevol}). 
We ran a sufficiently large number of impurity configurations 
(typically 40) to make the statistical error negligible \cite{note-statistics}.
The resulting error bars arise from the quality of the fittings to the exponential 
decay of the spin survival. 
In agreement with our analytical results, an inverse quadratic 
dependence of $\tau_\mathrm{s}$ on the coupling strength is obtained 
(Fig.\ \ref{fig:dynamevol}). 
Fitting this dependence of $\tau_\mathrm{s}$ on $\eta$ allows us to extract 
the physical values ($\eta=1$) of $\tau_\mathrm{s}$. 

In Fig.~\ref{fig:relaxtime} we present the spin relaxation times resulting from 
our numerical approach for GaAs at four different impurity densities above the MIT 
(black dots), together with the prediction of Eq.\ \eqref{eq:taus_vs_dens} 
(solid line), and 
the available experimental data from Refs.\ 
\cite{kik-aws,oes-roe-hau-hae,dzh,roe-ber-mue}. 
The agreement between the analytical and numerical approaches is 
very important in view of the complementarity of these two very different 
ways to extract $\tau_\mathrm{s}$. Both approaches describe the data 
within the experimental uncertainty and correctly reproduce the 
density dependence of the spin relaxation time. The small departure of 
the analytical and numerical results is not significant, taking into 
account the different approximations of both paths and the arbitrariness 
associated with the definition of relaxation times, i.e.\ numerical prefactors in
Eqs.\ \eqref{eq:reltimedef} and \eqref{eq:tauc}. 

While in the critical region and deep into the localized regime there is some dispersion 
of the experimental values for GaAs, depending on the different samples and measurement 
technique, on the metallic side of the MIT values of 
$\tau_\mathrm{s} \gtrsim \unit[100]{ns}$ are consistently obtained.
A decrease of $\tau_\mathrm{s}$ with $n_\mathrm{i}^{1/2}$ is observed, with a clear 
change in the density-dependence once the impurity and the conduction bands hybridize. 
Our analytical and numerical results of Fig.\ \ref{fig:relaxtime} (solid line and filled 
symbols) are obtained using the values $V_0=\unit[11.76]{meV}$ and 
$\gamma=\unit[27]{eV\AA^3}$ without any adjustable parameter. 
We remark that these 
results are very sensitive to the value of $\gamma$. For instance, taking 
$\gamma=\unit[15]{eV\AA^3}$, that is proposed by some measurements 
\cite{fab-etal,jusserand}, results in the dashed line. 
The identification of the Dresselhaus coupling as the dominant channel for spin 
relaxation close to the MIT of 3-dimensional samples will help in the precise 
determination of the material constant $\gamma$, which is crucial 
for the operation of low-dimensional spintronic devices. 

In conclusion, we have identified a spin relaxation mechanism characteristic of
electrons on the metallic side of the metal-insulator transition in the impurity 
band of semiconductors, where up to now a theoretical understanding was lacking, 
thereby solving a longstanding problem in spintronics.
Our mechanism is derived from the Dresselhaus spin-orbit coupling. It 
dominates over the usually stronger Rashba coupling in the landscape of 
hydrogenic impurities in semiconductors with zincblende structure, 
and provides relaxation times that agree with the experimentally measured values.

We gratefully acknowledge support from the ANR through grant 
ANR-08-BLAN-0030-02, the Coll\`{e}ge Doctoral Europ\'{e}en of Strasbourg, 
UBACYT (grant X495), ANPCYT (grant PICT 2006-02134), 
and from the program ECOS-Sud (action A10E06).


\begin{thebibliography}{99}
\bibitem{note-tauc}
We have obtained an excellent agreement between Eq.\ \eqref{eq:tauc} and 
the numerically computed time-evolution of an initially localized state within the MT model.
\bibitem{note-statistics}
The $\eta$- and $N$-dependent values of $\tau_\mathrm{s}$ do not depend on 
whether we extract them from the impurity-averaged spin survival curve or 
by averaging the rates of different configurations.
\end{thebibliography}
\end{document}